\documentclass[12pt]{article}
\usepackage{times,graphicx,theorem,epsfig,lscape,amsmath,dcolumn,multirow,epic,float,enumerate,comment}
\usepackage{natbib}
\usepackage{amsmath,amssymb}
\usepackage[normalem]{ulem}

\newcommand{\blinding}[2]{#1}   

\renewcommand{\baselinestretch}{1.60}

\theoremstyle{plain} 

\theoremstyle{plain}

\theoremstyle{plain} 
\theoremstyle{plain}

\theoremstyle{plain}

\newcommand{\one}{\mathbf{1}}

\DeclareMathOperator\bE{\mathbb E} 
\DeclareMathOperator\bV{\mathbb V} 

\newcommand{\bX}{\mathbf{X}}
\newcommand{\bZ}{\mathbf{Z}}

\newcommand{\ATE}{\mbox{\tiny{ATE}}}

\newcommand{\ATT}{\mbox{\tiny{ATT}}}
\newcommand{\ATC}{\mbox{\tiny{ATC}}}

\newcommand{\eh}{{\hat{e}_i}}

\begin{document}

\begin{center}
\vspace*{-2.5cm}

{\Large Balancing Covariates via Propensity Score Weighting}

\medskip
\blinding{
Fan Li \quad \quad Kari Lock Morgan \quad \quad Alan M. Zaslavsky
\footnote{Fan Li is associate professor, Department of Statistical Science, Duke University, Durham, NC, 27705 (email: fli@stat.duke.edu); Kari Lock Morgan is assistant professor, Department of Statistics, Penn State University, University Park, PA 16802 (email: klm47@psu.edu); Alan M. Zaslavsky is professor, Department of Health Care Policy, Harvard Medical School, Boston, MA 02115  (email: zaslavsk@hcp.med.harvard.edu). The authors are grateful to the associate editor and two anonymous reviewers for comments that help improve the clarity and exposition of the article, to Peng Ding for insightful discussions, particularly the proof of Corollary 1, to Dylan Small for sharing the programming code for the RHC study, and to Maggie Nguyen for computational assistance. Li and Morgan's research is partially funded by NSF-SES grant 1424688.}

}{}

\end{center}

\date{}

{\centerline{ABSTRACT}

\noindent Covariate balance is crucial for unconfounded descriptive or causal comparisons. However, lack of balance is common in observational studies. This article considers weighting strategies for balancing covariates. We define a general class of weights---the balancing weights---that balance the weighted distributions of the covariates between treatment groups.  These weights incorporate the propensity score to weight each group to an analyst-selected target population.  This class unifies existing weighting methods, including commonly used weights such as inverse-probability weights as special cases. General large-sample results on nonparametric estimation based on these weights are derived. We further propose a new weighting scheme, the overlap weights, in which each unit's weight is proportional to the probability of that unit being assigned to the opposite group. The overlap weights are bounded, and minimize the asymptotic variance of the weighted average treatment effect among the class of balancing weights. The overlap weights also possess a desirable small-sample exact balance property, based on which we propose a new method that achieves exact balance for means of any selected set of covariates. Two applications illustrate these methods and compare them with other approaches.

\vspace*{0.3cm}
}

\clearpage

\section{Introduction}
Unconfounded comparison between groups is a central goal in many observational studies. Comparative effectiveness studies aim to estimate the causal effect of a treatment or intervention unconfounded by differences between characteristics of those assigned to the treatment and control conditions under current practice. Noncausal descriptive studies often concern a controlled and unconfounded comparison of two populations, such as comparing outcomes among populations of different races or of cohorts in different years while giving the comparison groups similar distributions of some important covariates. Whether the purpose of the study is causal or descriptive, comparisons between groups can be biased when the groups lack balance, that is, have substantially different distributions of relevant covariates.

Standard parametric adjustment by regression is often sensitive to model misspecification when groups differ greatly in observed characteristics \citep{Rubin79}. A commonly used nonparametric balancing strategy is weighting, which applies weights to the sample of units in each treatment group to match the covariate distribution of a target population, and the comparison is made between the weighted outcomes. The literature on weighting \citep[e.g.][]{RobinsRotnitzky95,Hahn98,Robins00, Hirano01, Hirano03, Imbens04, Crump09} has been dominated by the inverse-probability weights (IPW), originating from survey research. A special case of IPW is the Horvitz-Thompson (HT) weight \citep{HT52}, which for each unit is the inverse of the probability of that unit being assigned to the observed group. The HT weights correspond to estimands defined on the population represented by the combined treatment and control groups, such as the average treatment effect (ATE) in causal studies. Variants of IPWs targeting at treatment effects on the treated group (ATT) or other subpopulations have also been commonly used \citep[e.g.][]{Hirano01, LiGreene13}.

The ATE and ATT estimands are widely accepted, largely because they represent effects on easy-to-interpret target populations, and estimators of these effects possess desirable statistical properties. For example, \cite{Manski90} showed that in cases where the outcome variable is bounded, sharp bounds on the ATE and ATT can be derived. However the automatic focus on these estimands is often questionable in practice. First, ATE and ATT may correspond to the effect of an infeasible intervention. For example, the ATE corresponds to the effect of switching every unit in the study population from one treatment to the other; such a complete change in treatment is rarely conceivable in medical studies, where the treatment might be known to be harmful to patients with certain characteristics. \cite{Imbens09} made a similar argument in the context of a job training program.  Second, as \cite{Rosenbaum12} explains, ``\dots often the available data do not represent a natural population, and so there is no compelling reason to estimate the effect of the treatment on all people recorded in this source of data\dots'' In such cases, applying the HT or ATT weights to the observed sample might not yield an estimate of the average treatment effect in the scientifically appropriate target population. Third, researchers commonly exclude some units from the final analysis, e.g. unmatched units or units with extremely large weights. The remaining sample may only represent a subpopulation of the original targeted population and this subpopulation can vary from study to study. Fourth, with ATE and ATT extreme probabilities (close to 0 or 1) introduce extreme IPWs, with adverse finite-sample consequences such as poor balance and large variance \citep{Busso14}. Lastly, in causal inference there is ample precedent for prioritizing covariate balance between treatment groups over generalizability to a recognizable target population, as demonstrated by the huge body of experimental evidence based on randomization of treatment within a non-randomly selected (in many cases convenience) sample.

A recent strand of literature suggests shifting the focus from ATE or ATT to subpopulations with sufficient overlap in covariates between treatments. \cite{Crump09} advocated focusing on the ``optimal'' subpopulation for which the average treatment effect can be estimated with the smallest variance, approximated by discarding all units with estimated propensity scores outside the range $[\alpha, 1-\alpha]$. However, as illustrated in Section~\ref{sec:disparity}, this approach can be sensitive to the choice of the cutoff point, for which there is no strong theoretical guidance, and can possibly exclude much of the study sample.  In an example in \cite{Crump06}, setting $\alpha$ to 0.1 removes 90\% of the study sample; the same will occur if one of the groups is much larger than the other, driving propensity scores close to 0 or 1. Moreover, the propensity-score-based cutoff criterion may be difficult to interpret in practice \citep{TraskinSmall11}.

In this article, we define the class of \emph{balancing weights} that balance the distributions of covariates between comparison groups for any pre-specified target population. We stress that the choice of target population distribution is neither automatic nor restricted to the ATT and ATE. Within this class of weights, we introduce the \emph{overlap weights}, which weight each unit proportional to its probability of assignment to the \emph{opposite} group. Unlike the IPWs, the overlap weights are bounded between 0 and 1 so the weighted distributions are always integrable. Under mild conditions, the overlap weights minimize the asymptotic variance of the nonparametric estimate of the weighted average treatment effect within the class of balancing weights.

Applications of these weights have appeared in the health services literature at least since 2001 \blinding{\citep{Schneider01}}{[blinded citation]} but without theoretical development.  Large-sample results on overlap weights appeared in the working paper of \cite{Crump06}, who regarded the subpopulation defined by the overlap weights as merely a device to ``acknowledge and address the difficulties in making inferences about the population of primary interest.'' In contrast, we foreground the overlap weights as defining a population of substantial clinical or policy interest, namely the subpopulation which currently receives either treatment in substantial proportions. We illustrate the substantive relevance of this target population and estimand in two applications, one concerning racial disparities in medical expenditure and one concerning evaluation of a medical treatment. Overlap weights estimated from a logistic model also have a useful small-sample property: they yield \emph{exact balance} between groups in the means of each covariate included in the model. This property can be exploited in a hybrid  that imposes exact balance on a matched sample by reweighting it.

Section \ref{sec:fundamentals} introduces the general framework of balancing weights and defines corresponding estimands. Section \ref{sec:largesample} presents three large-sample properties of the nonparametric weighting estimators.  In Section \ref{sec:overlap} we propose the overlap weights, present a small-sample exact balance result, and describe the new matching-weighting hybrid method. In Section \ref{sec:examples}, we illustrate the proposed methods through simulations and two real applications. Section \ref{sec:conclusion} concludes.

\section{Balancing Weights} \label{sec:fundamentals}
Consider a sample or finite population of $n$ units, each belonging to one of two groups for which covariate-balanced comparisons are of interest, possibly defined by a treatment.  Let $Z_{i}=z$ be the binary variable indicating membership in groups that may be labeled treatment ($z=1$) and control ($z=0$). For each unit, an outcome $Y_i$ and a set of covariates $X_i=(X_{i1},...,X_{iK})$ are observed. The propensity score \citep{Rosenbaum83} is the probability of assignment to the treatment group given the covariates,  $e(x)=\Pr(Z_i=1|X_i=x)$.

In descriptive comparisons, ``assignment'' is to a nonmanipulable state defining membership in one of two groups, and a common objective is to evaluate the average difference in the outcome in two groups with balanced distributions of covariates. We first define the conditional average controlled difference (ACD) \citep{LiLandrumZaslavsky13} for a given $x$ as,
\begin{equation}
\tau(x)\equiv \bE(Y|Z=1,X=x)-\bE(Y|Z=0,X=x). \label{ACD}
\end{equation}

For causal comparisons, we adopt the potential outcome framework \citep{Rubin74,Rubin78,ImbensRubin15}. Assuming the standard Stable Unit Treatment Value Assumption (SUTVA) \citep{Rubin80}, which states that the potential outcomes for each unit are unaffected by the treatment assignments of other units, each unit has potential outcomes $\{Y_{i}(z), z=0,1\}$ corresponding to the possible treatment levels, of which only one is observed: $Y_i=Z_iY_i(1)+(1-Z_i)Y_i(0)$.  Under the \emph{unconfoundedness} assumption, that is, $\{Y(0), Y(1)\} \perp Z |X$, we have $\Pr(Y(z)|X)=\Pr(Y|X,Z=z)$ for $z=0,1$, so $\tau(x)$ in \eqref{ACD} is a causal estimand---the average treatment effect (ATE) conditional on $x$:
\begin{equation}
\tau(x)= \bE[Y(1)-Y(0)|X=x].
\end{equation}
Estimation of either comparison requires the \emph{probabilistic assignment assumption}, $0<e(X)<1$, which states that the study population is restricted to values of covariates for which there can be both control and treated units. Assignment mechanisms assuming both probabilistic assignment and unconfoundedness are \emph{strongly ignorable} \citep{Rosenbaum83}.

Typically in either descriptive or causal comparisons the (potential) outcomes are compared not for a single $x$ but rather averaged over a hypothesized target distribution of the covariates. Assume the marginal density of the covariates $X$, $f(x)$, exists, with respect to a base measure $\mu$ (a product of counting measure with respect to categorical variables and Lebesgue measure for continuous variables).  We then represent the target population density by $f(x)h(x)$, where $h(\cdot)$ is pre-specified function of $x$, and define a general class of estimands by the expectation of the conditional ACD or ATE $\tau(x)$ over the target population:
\begin{equation}
\tau_{h} \equiv \frac{\int \tau(dx)f(x)h(x)\mu(dx)}{\int f(x)h(x)\mu(dx)}. \label{wate}
\end{equation}
For descriptive comparisons, $\tau_h$ is the weighted ACD, and for causal comparisons it is the weighted average treatment effect (WATE) \citep{Hirano03}, the term we use henceforth for either case.

Let $f_z(x)=\Pr(X=x|Z=z)$ be the density of $X$ in the $Z=z$ group, then
\[f_1(x)\propto f(x)e(x), \quad  \mbox{and} \quad f_0(x)\propto f(x)(1-e(x)).\]
For a given $h(x)$, to estimate $\tau_h$, we can weight $f_z(x)$ to the target population using the following weights (proportional up to a normalizing constant):
\begin{equation}
\left\{
\begin{array}{ll}
w_1(x) \propto  \frac{f(x)h(x)}{f(x)e(x)} =\frac{h(x)}{e(x)}, &\mbox{for } Z=1, \\
w_0(x) \propto  \frac{f(x)h(x)}{f(x)(1-e(x))} =\frac{h(x)}{1-e(x)}, & \mbox{for } Z=0.
\end{array}
\right. \label{weight}
\end{equation}
We call this class of weights defined in \eqref{weight} the \emph{balancing weights} because they balance the weighted distributions of the covariates between comparison groups:
\begin{equation}
f_1(x)w_1(x) = f_0(x)w_0(x) = f(x)h(x).
\end{equation}
The function $h$ can take any form, and all weights that balance the covariate distributions between groups can be specified within this class.

Specification of $h$ defines the target population and estimands and determines the weights. Statistical, scientific and policy considerations all may come into play. When $h(x)=1$, the corresponding target population $f(x)$ is the combined (treated and control) population, the weights ($w_1, w_0$) are the HT weights $(1/e(x), 1/(1-e(x)))$, and the estimand is the ATE for the combined population, $\tau^{\ATE}=\bE[Y(1)-Y(0)]$. When $h(x)=e(x)$, the target population is the treated subpopulation, the weights are $(1, e(x)/(1-e(x)))$, and the estimand is the average treatment effect for the treated (ATT), $\tau^{\ATT}=\bE[Y(1)-Y(0)|Z=1]$. When $h(x)=1-e(x)$, the target population is the control subpopulation, the weights are $((1-e(x))/e(x), 1)$, and the estimand is the average treatment effect for the control (ATC), $\tau^{\ATC}=\bE[Y(1)-Y(0)|Z=0]$. By choosing $h$ from the class of indicator functions, one can define the ATE for truncated subpopulations of substantive interest or with desirable theoretical properties. For example, \cite{Crump09} recommended use of $h(x)=\one(\alpha<e(x)<1-\alpha)$ with a pre-specified $\alpha \in (0,1/2)$ that defines a subpopulation with sufficient overlap of covariates between two groups. \cite{LiGreene13} proposed to use $h(x)=\min\{e(x), 1-e(x)\}$, which defines a weighting analogue to pair matching; a similar notion was discussed earlier in \cite{Dehejia99}. These examples are summarized in Table \ref{tab:weight}.  Formulation \eqref{weight} also  allows application-specific choices of $h$ that are not functions of the propensity score, possibly to select a subpopulation of interest (like an age range) or to match its covariate distribution.
\begin{table}
\begin{center}
\begin{tabular}{c|c|c|c}
\hline
target population  &$h(x)$ & estimand & weight $(w_1, w_0)$  \\
\hline
combined &1         &ATE   &$\left(\frac{1}{e(x)}, \frac{1}{1-e(x)}\right)$ [HT]   \\
treated  &$e(x)$    &ATT   &$\left(1, \frac{e(x)}{1-e(x)}\right)$   \\
control  &$1-e(x)$  &ATC   &$\left(\frac{1-e(x)}{e(x)}, 1\right)$   \\
overlap  &$e(x)(1-e(x))$ &ATO & $(1-e(x), e(x))$  \\
truncated combined  &$\one(\alpha<e(x)<1-\alpha)$ &  &$\left(\frac{\one(\alpha<e(x)<1-\alpha)}{e(x)}, \frac{\one(\alpha<e(x)<1-\alpha)}{1-e(x)}\right)$  \\
matching & $\min\{e(x), 1-e(x)\}$ &&  $\left(\frac{\min\{e(x), 1-e(x)\}}{e(x)}, \frac{\min\{e(x), 1-e(x)\}}{1-e(x)}\right)$\\
\hline
\end{tabular}
\caption{Examples of balancing weights and corresponding target population and estimand under different $h$.} \label{tab:weight}
\end{center}
\end{table}

\section{Large-sample Properties of Nonparametric Estimators} \label{sec:largesample}

Here we establish properties of the \emph{sample estimator of WATE},
\begin{equation}
\label{eq:sampleWATE}
\hat{\tau}_h=\frac{\sum_i w_1(x_i)Z_i Y_i}{\sum_i w_1(x_i)Z_i} -
              \frac{\sum_i w_0(x_i)(1-Z_i) Y_i}{\sum_i w_0(x_i)(1-Z_i)}
\end{equation}
where the sum is over a sample drawn from density $f(x)$.
Proofs of the following three large-sample results regarding $\hat{\tau}_h$ are in the Appendix; a rigorous development of these and other properties of nonparametric estimators under the required regularity conditions appears in \cite{Imbens04} and \cite{Crump06}.

\textbf{Theorem 1.} \emph{$\hat{\tau}_h$ is a consistent estimator of $\tau_h$.}

The next result concerns the component of variation due to residual (model) variation in $\hat{\tau}_h$ conditional on the sampled covariate design points $\bX=\{x_1,\dots,x_n\}$, the first term of the decomposition $\bV[\hat{\tau}_h]=\bE_x\bV[\hat{\tau}_h\mid \bX]+\bV_{x}\bE[\hat{\tau}_h\mid \bX]$,  showing that it can be characterized after making only limited assumptions about residual variances, as in the Corollary below. The second term arises from the dependence of the expectation of \eqref{eq:sampleWATE} on the sample, and estimating it involves the outcome model (associations between $Y(z)$ and $x$).  However, as \cite{Imbens04} argues, individual variation is typically much larger than conditional mean variation, so the benefit of further optimizing the weights by a preliminary look at the outcomes typically would not justify the risk of biasing model specification to attain desired results.  Hence our weighting strategy is based on general results concerning the first term.

\textbf{Theorem 2.} \emph{As $n\rightarrow\infty$, the expectation (over possible samples of covariate values) of the conditional variance of the estimator $\hat{\tau}_h$ given the sample $\bX=\{x_1,\dots,x_n\}$ converges:}
\begin{eqnarray}
n\cdot\bE_x\bV[\hat{\tau}_h\mid\bX] &\rightarrow&\int f(x)h(x)^2\left[v_1(x)/e(x)+v_0(x)/(1-e(x))\right] \mu(dx)\big/C_h^2, \label{variance}
\end{eqnarray}
\emph{where $v_z(x)=\bV[Y(z)\mid\bX]$ and $C_h=\int h(x)f(x)d\mu(x)$ is a normalizing constant.}

Consequently, if the residual variance is assumed to be homoscedastic across both groups, $v_1(x)=v_0(x)=v$, then the asymptotic variance of $\hat{\tau}_h$ simplifies to
\begin{equation}\label{simplevar}
n\cdot\bE_x\bV[\hat{\tau}_h\mid\bX] \rightarrow v/C_h^2 \int \frac{f(x)h(x)^2\mu(dx)}{e(x)(1-e(x))}.
\end{equation}

\textbf{Corollary 1.}  \emph{The function $h(x)\propto e(x)(1-e(x))$ gives the smallest asymptotic variance for the weighted estimator $\hat{\tau}_h$ among all $h$'s under homoscedasticity, and as $n\rightarrow\infty$,}
\begin{equation*}
n\cdot\mbox{min}_h\{\bE_x\bV[\hat{\tau}_h\mid\bX]\} \rightarrow v/C_h^2 \int f(x)e(x)(1-e(x))\mu(dx).
\end{equation*}

In applications, the true propensity score $e$ is unknown and is replaced by the estimated propensity score $\hat{e}$. As shown in \cite{Rosenbaum87} and \cite{Hirano03}, a consistent estimate of the propensity score in fact leads to more efficient estimation than the true propensity score.

These theoretical results have important practical consequences for propensity score weighting analysis. First, calibration of the propensity score model is important:  the predicted and empirical rates of treatment assignment should agree in relevant subsets of the covariate space, or else covariate balance cannot be attained.  A simple check is to compare predicted and observed rates in deciles (or other convenient intervals appropriate to the amount of data) of the score distribution, in the spirit of \cite{hosmer:lemeshow}.  If miscalibration is identified, likely indicating a misfit in the link function, it can be corrected by ad hoc methods such as ratio adjustment of each decile or adding indicator variables for the deciles of a trial model to the final model (if the error is relatively small), or by fitting a flexible logistic spline model relating the estimated linear predictor to treatment assignment. Estimates of treatment effect by decile may also be of scientific interest for assessing association of treatment {\em assignment} with treatment {\em effect}.

Second, a rich propensity score model, rather than a parsimonious one, is desirable, especially in causal inference applications, because the ignorability assumptions are likely to be violated if the propensity score is only a simple logistic-linear function of the covariates but both the outcome and assignment mechanism are functions of more complex interactions or nonlinear terms.  Given this, traditional statistical tests of goodness of fit are only minimally relevant, because the objective of weighting is to balance covariate distributions in  the {\em sample}, not to make inferences about assignment probabilities in the {\em population}.  Rather, the limitation on model complexity is imposed by a bias-variance tradeoff.  As the model becomes more complex and therefore more predictive, propensity scores move toward 0 and 1, becoming exactly 0 or 1 when the model discovers a separating plane in the data. Thus the weights $h(x)=e(x)(1-e(x))$ for many design points move toward zero, reducing the precision of estimates.  Analytic judgment is required to decide when further potential reductions in  bias are outweighed by the cost in variance.  Nonetheless, maintaining the principle of separating design (the weighting model) from analysis (using outcome data), the variance inflation due to model complexity can be approximated by calculating the corresponding ratio for an estimated difference of weighted means assuming homoscedastic data; from \eqref{simplevar} this is estimated by
\begin{equation}
(1/n_1+1/n_0)^{-1}\sum_{z=0,1}\left(\sum_{i=1}^{n_z}w_{zi}^2\right)\Big/ \left(\sum_{i=1}^{n_z}w_{zi}\right)^2, \label{varianceinflation}
\end{equation}
based on the ``design effect" approximation of \cite{kish}.

\section{Overlap Weighting} \label{sec:overlap}
\subsection{The Overlap Weights}
As is generally recognized, ATE is inestimable if the supports of the treatment groups are different, making balanced comparisons impossible in the region of support for only one group.  Furthermore, the value and indeed the finiteness of the variance in (\ref{variance}) depends critically on the distribution of $e(x)$ close to 0 and 1 unless the denominator $e(x)\cdot\left(1-e(x)\right)$ is canceled by a corresponding factor of $h(x)$.

We define the \emph{overlap weights} by letting $h(x)=e(x)(1-e(x))$, implying balancing weights
\begin{equation}
\left\{
\begin{array}{ll}
w_1(x) \propto 1-e(x), &\mbox{for } Z=1, \\
w_0(x) \propto e(x),  & \mbox{for } Z=0.
\end{array}
\right.
\end{equation}
Following Corollary 1 and under its assumptions, the corresponding nonparametric estimator $\hat{\tau}_h$ has the minimum asymptotic variance among all balancing weights. Even if the outcomes are moderately heteroscedastic, the overlap weights may have good efficiency while depending only on design (covariate distributions in the treatment and control groups) and therefore being defined prior to examining outcome data.

\begin{figure}[!ht]
	\centering\includegraphics[width=.75\textwidth]{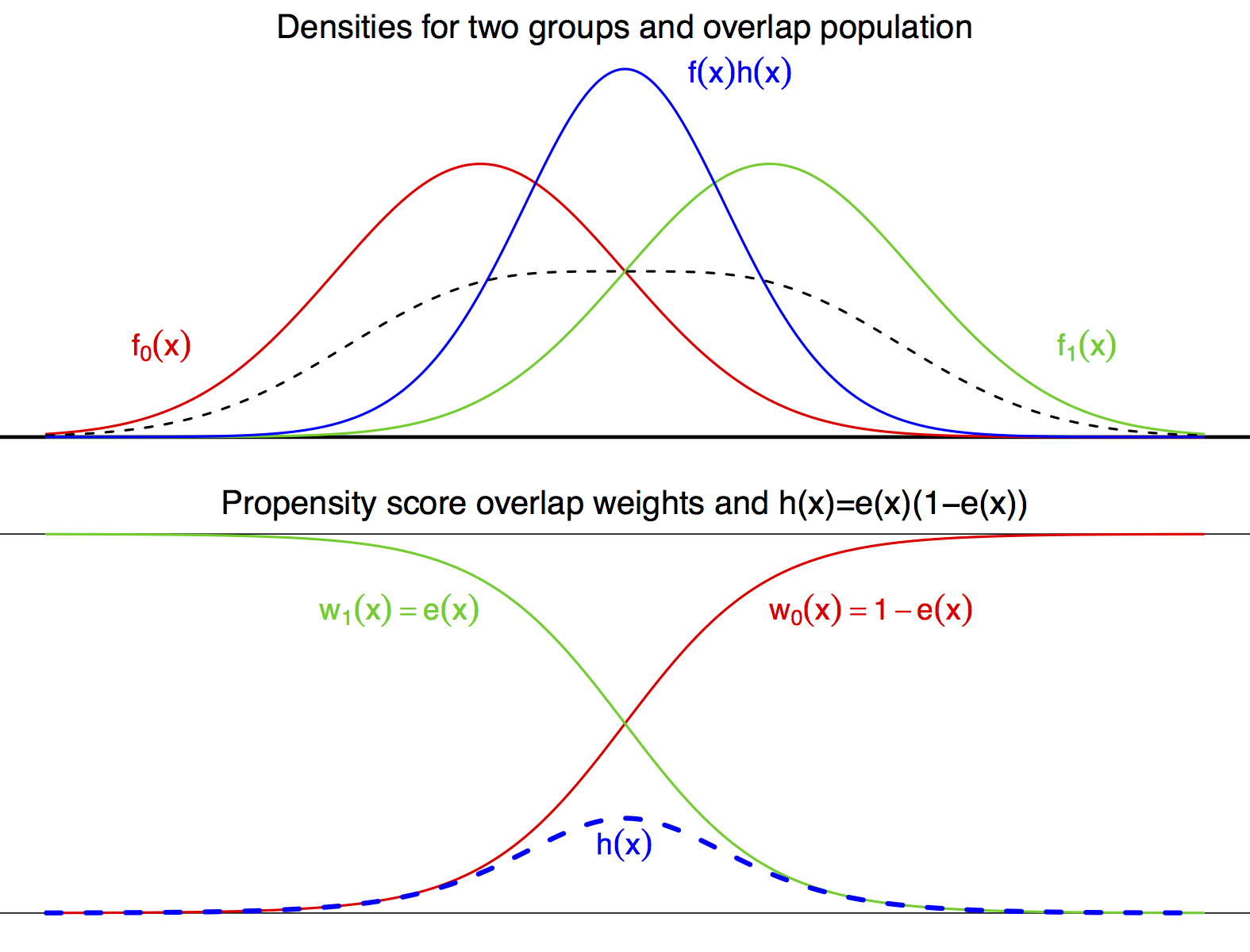}
	\label{fig:overlap}
	\caption{Overlap weights for two normally-distributed groups with different means. In the upper panel, the left and right solid lines, the thin and thick dashed lines represent the density of the covariate in the control, treated, combined $(h(x)=1)$, and overlap weighted ($h(x)=e(x)(1-e(x))$) populations, respectively. In the lower panel, the two solid lines represent $w_0(x)$, $w_1(x)$ and the dashed line represents $h(x)=e(x)(1-e(x))$.}
\end{figure}

The overlap weights and the associated target population are illustrated in Figure \ref{fig:overlap} for two univariate normal populations with equal size and variance. The upper panel illustrates the target population density $f(x)h(x)$, which is greatest where the treated and control groups most overlap. Hence, we call the corresponding WATE estimand $\tau_h$ the \emph{average treatment effect for the overlap population} (ATO). The lower panel shows that the ratio $h(x)$ of target to combined population peaks where the propensity score is 1/2, and these weights place more emphasis on units with propensity score close to 1/2, who could be in either group, relative to those with propensity scores close to 0 or 1.

The overlap weights, unlike the IPWs, are bounded and thus are less sensitive to extreme weights. Compared to the common practice of truncating weights or discarding units, the overlap weights are continuously defined and avoid arbitrary choice of a cutoff for inclusion in the analysis, as illustrated in an application in Section \ref{sec:disparity}.

The variance-minimizing property of the overlap weighting implies that it adapts to any distribution of covariates and propensities to define and answer the question that can be best answered nonparametrically by the data at hand.  These include several familiar special cases: (1) if propensity to treatment is always small, ATO approximates ATT (for $e(x)\approx 0$, $(1-e(x),e(x))\approx\left(1,\frac{e(x)}{1-e(x)}\right)$); (2) in the opposite case where propensity to control is small, ATO approximates ATC; (3) if treatment and control groups are nearly balanced in size and distribution, ATO approximates ATE (for $e(x)\approx 1/2$, $(1-e(x),e(x))\approx\left(\frac{.25}{e(x)},\frac{.25}{1-e(x)}\right)$).

While the interpretation of the overlap distribution is specific to the application, the overlap population often represents a target population of intrinsic substantive interest, that is, the units whose combination of characteristics could appear with substantial probability in either treatment group.  For example, in medicine, these may be patients for whom clinical consensus is ambiguous or divided, who are said to be in equipoise between treatments, so research on these patients may be most needed. In social policy, these might be the units whose treatment assignment would be most responsive to a policy shift as new information is obtained.  In general, we may want to estimate a treatment effect or compare treatments on the subpopulation of units that are the most similar between treatment groups.

Overlap weighting is in a sense asymptotically equivalent to matching. Consider a sequence of increasingly large datasets from some generating distribution.  The weighting analyses might be a sequence of increasingly complex models that converge to a saturated propensity score model with indicators for each design point (or small neighborhood, for continuous variables), while the matching criterion of closeness is correspondingly tightened to approach exact matching on the same discrete design points or continuous neighborhoods. At this limit, many-to-many matching would use weights equivalent to the overlap weights estimated from the saturated propensity score model. The overlap weights also have a natural connection to regression with fixed effects for each design point. If the sample count for $x_i$ in group $z=0,1$ is $n_{zi}$, the propensity score is $e(x_i)=n_{1i}/(n_{0i}+n_{1i})$ and the total overlap weight for each group and hence of $\hat{\tau}(x_i)$ is $n_{0i}n_{1i}/(n_{0i}+n_{1i})$; however this is exactly the precision weight attached to $\bar{Y}_{1i}-\bar{Y}_{0i}$ in the fixed-effects OLS model $Y_{zi}=\alpha_{i}+z\tau+\epsilon_{zi}$.

\subsection{Exact Balance}
Overlap weights based on a logistic propensity score model also have the following attractive small-sample exact balance property (proved in the Appendix).

\textbf{Theorem 3.} \emph{When the propensity scores are estimated by maximum likelihood under a logistic regression model, $\mathrm{logit}\;e(x_i)=\beta_0+x_i\beta'$, the overlap weights lead to exact balance in the means of any included covariate between treatment and control groups. That is, }
\begin{equation} \label{eq:exact_balance}
\frac{\sum_i x_{ik}Z_i(1-\hat{e}_i)}{\sum_i Z_i(1-\hat{e}_i)}
=\frac{\sum_i x_{ik}(1-Z_i)\hat{e}_i}{\sum_i (1-Z_i)\hat{e}_i}, \quad \mbox{for} ~~ k=1,\dots, K,
\end{equation}
\emph{where $\hat{e}_i= \{1+\exp[-(\hat{\beta}_0 +x_i\hat{\beta}')]\}^{-1}$ and $\hat{\beta}=(\hat{\beta}_1,...,\hat{\beta}_k)$ is the MLE for the regression coefficients}.

While a main effects model guarantees exact equality between groups for the mean of each included covariate, it is advisable to improve balance by including additional derived covariates, guided by prior anticipation of possible effects on outcomes, as discussed at the end of Section \ref{sec:largesample}.  These may include interactions and, for continuous variables, terms whose mean balance implies better matching of distributions, such as powers and products (to enforce equality of moments)  or spline terms. Note that the exact balance property applies for overlap weights with propensity scores calculated with the canonical link under any generalized linear model, including those for multi-group comparisons, e.g. multinomial logistic models for unordered categorical outcomes or nested logistic models for ordinal outcomes.

Theorem 3 is distinct from the exact balance result in \cite{Graham12}, which is on inverse probability weights and does not provide the explicit form of the propensity score model that would guarantee exact balance.

\subsection{Combining Matching and Weighting: The ``Tudor Solution"}  \label{sec:Tudor}
Matching is the most widely used nonparametric adjustment method in practice. Based on the exact balance property of the overlap weights (Theorem 3), we propose a hybrid approach that combines matching and weighting --- the ``Tudor solution:" \footnote{The name ``Tudor'' evokes the Wars of the Roses (1455-1485), in which the houses of Lancaster and York fought bitterly for the throne of England but victory went to a relatively remote Lancastrian claimant, the House of Tudor, which through a strategic marriage ultimately united the claims of the rival houses. Here it refers to the contention between weighting and matching in the statistical literature.} matching followed by an overlap weighting adjustment of the matched sample. In the first step, a matched sample is created using any preferred approach (e.g., Mahalanobis distance matching within propensity score calipers).  In the second step, propensity scores are estimated by logistic regression \emph{within} the matched sample. Finally the treatment effect is estimated by applying the overlap weights to the matched sample. This approach has the potential to combine the benefits of matching (nearness of matched cases in multivariate space, including dimensions not controlled by the propensity score model) and overlap weighting (exact balance for means of covariates in model). An analogous use of substitution sampling followed by regression adjustment in surveys was proposed by \cite{RubinZanutto02}.

The Tudor solution might be most beneficial for high-dimensional data, where considerable residual imbalance is possible due to its typical local sparseness. Removing mean imbalance also removes the component of bias that would be estimated by a post-matching linear model, but has the advantage of being conducted in the ``design phase," without knowledge and possible manipulation of consequences for the estimates \citep{Rubin08}.

\section{Examples}\label{sec:examples}
\subsection{Artificial Distributions}
It is easy to construct examples in which the asymptotic variance \eqref{variance} of estimates using HT weights is infinite. For example, let $x\sim\textrm{Uniform}(0,1)$, so  $f(x)\equiv 1$ and let $e(x)=x$.
	
Here we illustratively compare HT, truncated HT (discarding units outside $.1<e<.9$), and overlap weighting estimators under more plausible assumed distributions by calculating variances of WATE estimates relative to the variance of the unweighted difference of means under homoscedastic errors, for selected univariate distributions $F_1,F_0$ (Table \ref{tab:relvar}). As required by theory, variances with overlap weights are always the smallest. In scenario (1), there is a modest shift of normal distributions; the HT estimator loses efficiency due to extreme weights in the tails, which are excluded by the truncated estimator.  With the larger shift between groups in scenario (2), the HT variance is greatly inflated, but the truncated HT weighting removes most of the extreme weights in the tails.  In scenario (3) one sample is much larger than the other, which skews the propensity score distribution.  This causes an excessive truncation of one tail of that distribution, increasing the variance of the truncated HT estimator.  Adaptive modification of truncation points might solve this problem, but no modifications are needed for the overlap weighting estimator. In scenario (4) one group has much larger variance, again inflating the HT variance, causing an explosion of weights in the tails of the narrower distribution.
	
	\begin{table}\label{tab:relvar}
		\renewcommand{\baselinestretch}{1.1}
		\normalsize
		\begin{center}
			\begin{tabular}{lcccrrr}
				&       &       &           & \multicolumn{3}{c}{Relative variance} \\ \cline{5-7}
				& $F_1$ & $F_0$ & $n_0/n_1$ & HT & HT(trunc) & Overlap \\ \hline
				(1) & $N(0,1)$ & $N(1,1)$ & 1 & 1.43 & 1.36 & 1.26 \\
				(2) & $N(0,1)$ & $N(2,1)$ & 1 & 11.81 & 2.88 & 2.22 \\
				(3) & $N(0,1)$ & $N(1,1)$ & 20 & 2.48 & 3.31 & 1.06 \\
				(4) & $N(0,1)$ & $N(0,20^2)$ & 1 & 50.02 & 4.55 & 3.16 \\
			\end{tabular}
			\caption{Variances of WATE estimators relative to variance of difference of unweighted means, under homoscedasticity and various covariate distributions.}
		\end{center}
		
	\end{table}

\subsection{A Causal Comparison:  Right Heart Catheterization} \label{sec:RHC}
Right heart catheterization (RHC) is a diagnostic procedure for directly measuring cardiac function in critically ill patients. Though useful for directing immediate and subsequent treatment, RHC can cause serious complications. In an influential study \cite{Connors96} used propensity score matching to study the effectiveness of right heart catheterization (RHC) with observational data from \cite{Murphy90}. The study collected data on 5735 hospitalized adult patients at five medical centers in the U.S., 2184 of them assigned to the treatment ($Z_i=1$), receipt of RHC within 24 hours of admission, and the remaining 3551 assigned to the control condition ($Z_i=0$). The outcome was survival at 30 days after admission. Based on information from a panel of experts, a rich set of variables potentially relating to the decision to use RHC was collected. \cite{Connors96} describes the study, which has been since intensively re-analyzed  \citep[e.g.][]{Hirano01,Crump09,TraskinSmall11,Rosenbaum12}. The dataset is publicly available on \texttt{http://biostat.mc.vanderbilt.edu/wiki/pub/Main/DataSets/rhc.html}.

The comparison in the RHC study is causal in the sense that the treatment---application of RHC---is manipulable \citep{Rubin86}. Among the 72 observed covariates (21 continuous, 25 binary, 26 dummy variables from breaking up 6 categorical covariates), distributions of several key covariates differed substantially between the control and treatment groups \cite[Table 2]{Hirano01}. For example, the treated group has a much higher average APACHE (Acute Physiology and Chronic Health Evaluation) score, signifying greater severity of disease at admission. The majority of the treated units have propensity scores larger than 0.5 and the majority of the control units have propensity scores smaller than 0.5 \cite[Figure 1]{Crump09}. Most previous analyses focus on the ATT, that is, the average causal effect of applying RHC for the patients who received RHC. In this study, as argued by \cite{Rosenbaum12}, it would also be of interest to estimate the effects of RHC for the ``marginal'' subjects, who might or might not have been treated. Such estimates provide useful information for assigning treatments for the population with no clear propensity to a group. Towards this goal, \cite{Rosenbaum12} proposed an optimal matching strategy to choose the matches with the most treated subjects that have adequate balance; \cite{Crump09} limits the weighting analysis to a subsample with the estimated propensity scores truncated to the interval $[0.1,0.9]$.

\begin{figure}[!ht]
\centering
\includegraphics[scale=0.6]{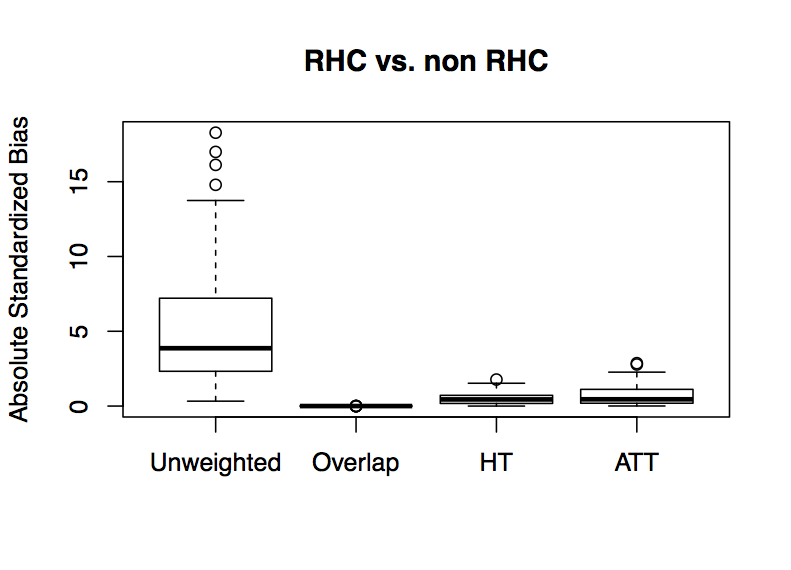}
\caption{Boxplots for the absolute standardized differences for covariates under each weighting method in the RHC study.}
\label{fig:box_rhc}
\end{figure}

As in previous studies, we estimate the propensity score under a logistic model with main effects of all the 72 covariates, based on which we calculate the HT, ATT and overlap weights. We measure covariate mean balance for each covariate by the absolute standardized bias (ASB):
\begin{equation}
\text{ASB} = {\left|\frac{\sum_{i=1}^N x_i Z_i w_i}{\sum_{i=1}^N Z_i w_i} - \frac{\sum_{i=1}^N x_i (1- Z_i) w_i}{\sum_{i=1}^N (1-Z_i) w_i}\right|}\Bigg /{\sqrt{s_{1}^2/N_1 + s_{0}^2/N_0}},\label{eq:std_bias}
\end{equation}
where $s_z^2$ is the variance of the unweighted covariate in group $z$ and $N_z$ is the sample size in group $z$.  For unweighted data, this is the absolute value of the standard two-sample t-statistic. We use an unweighted standard error in the denominator to allow for fair comparisons across weighting methods: if the numerator and denominator were both to vary with weighting method, a smaller ASB could be due to either better covariate balance or an increased standard error. We calculate the ASB for each covariate after applying the overlap, HT, and ATT weights, boxplots of which are displayed in Figure~\ref{fig:box_rhc}. Each set of weights improves mean balance compared to the unweighted data, but the overlap weights clearly lead to the best balance; in fact, here the ABS is exactly 0 for each covariate using overlap weights, based on Theorem 3.

The causal effects estimated from $\hat{\tau}_h$ using the three weights are shown in Table~\ref{tab:result_rhc}. For comparison, we include results from the method of \cite{Crump09} for ATT weighting with optimal truncation, and the optimal matching method of \cite{Rosenbaum12}. The standard errors are estimated from the bootstrap procedure in \cite{Crump09} with 5000 replicates. All the estimates suggest that applying RHC leads to a higher mortality rate than not applying RHC. Overlap weighting and optimal truncation lead to the smallest standard errors.  Overlap weighting and optimal matching give similar point estimates, around 10\% larger than those from the other methods.

\begin{table}[!ht]
\begin{center}
	\renewcommand{\baselinestretch}{1.1}
	\normalsize
\begin{tabular}{l|ccccccc}
\hline
                        & unweighted & overlap & HT & ATT & Trunc. ATT & Opt. match  \\
\hline
Estimate $\times 10^2$    & -7.36 & -6.54 & -5.93 & -5.81 & -5.90 & -6.78  \\
SE   $\times 10^2$        & 1.27 & 1.32 & 2.46 & 2.67 & 1.43& 1.56  \\
\hline
\end{tabular}
\caption{Estimates and standard errors obtained from different methods. Estimates using optimal truncation are from \cite{Crump09}, with estimated propensity score truncated between $[0.1, 0.9]$ (sample size 4728). Estimates using optimal matching  are from \cite{TraskinSmall11} based on 1563 optimally matched pairs.  Standard errors are calculated via bootstrapping.}
\label{tab:result_rhc}
\end{center}
\end{table}

To compare (truncated) weighting, matching and the proposed ``Tudor Solution", we conduct a simulation study closely mimicking the real RHC data. We retain the covariates and the treatment variable; we generate the outcome from a linear regression model with the treatment and main effects of the 72 covariates, with coefficients being the ones from fitting such a model to the real data. We simulate 1000 replicates of the data, each created by sampling with replacement 2184 and 3551 units from the treated and control group, respectively. We focus on the ATT. For weighting, only units with a propensity score between 0.1 and 0.9 are included. Absolute biases (in $10^{-2}$ scale) from the truncated weighting, matching and Tudor analyses are 1.10, 1.44 and 0.81, respectively, and root mean squared errors (RMSE) are 1.31, 1.75, and 1.01 respectively. The Tudor solution significantly outperforms matching alone, resulting in about 40\% reduction in both bias and RMSE. This large gain is not surprising given the high dimensional covariates and the underlying linear outcome model. The Tudor solution also outperforms truncated weighting, resulting in about 25\% reduction in both bias and RMSE.

\subsection{A Descriptive Comparison:  Racial Disparities in Medical Expenditure}\label{sec:disparity}
This analysis of racial disparities in medical expenditure uses data similar to that in  \cite{lecook10} for adult respondents aged 18 and older to the 2009 Medical Expenditure Panel Survey (MEPS); the dataset is publicly available on AHRQ's website \citep{meps09}. The sample contains 10,130 non-Hispanic Whites (referred to as Whites hereafter), 4224 Blacks, 1522 Asians, and 5558 Hispanics. The goal is to estimate disparities in health care expenditure between Whites and each minority group after balancing confounding variables. To illustrate, we focus on two separate comparisons -- the White-Asian and White-Hispanic comparisons. Since race is not manipulable, these comparisons are descriptive.

Here, focusing on ATT or ATE would force us to consider a hypothetical target population with the covariate distribution of a particular race or a ``combined" White-minority population. Even if such a  population is not an infeasible target of inference due to lack of overlap, it may lack policy relevance for studying disparities because it focuses on individuals atypical for their own racial/ethnic group. For example, body mass index (BMI) has a much longer and heavier right tail for Whites than for Asians, so weighting Asians to have a BMI distribution of the Whites creates an unrealistic population of Asians.  Instead, we choose the overlap weights with the goal of focusing on the naturally comparable subpopulation of people with similar characteristics: people who, based on their covariates, could easily be either White or from the minority group.

\begin{figure}[!ht]
\centering
\includegraphics[width=.9\textwidth]{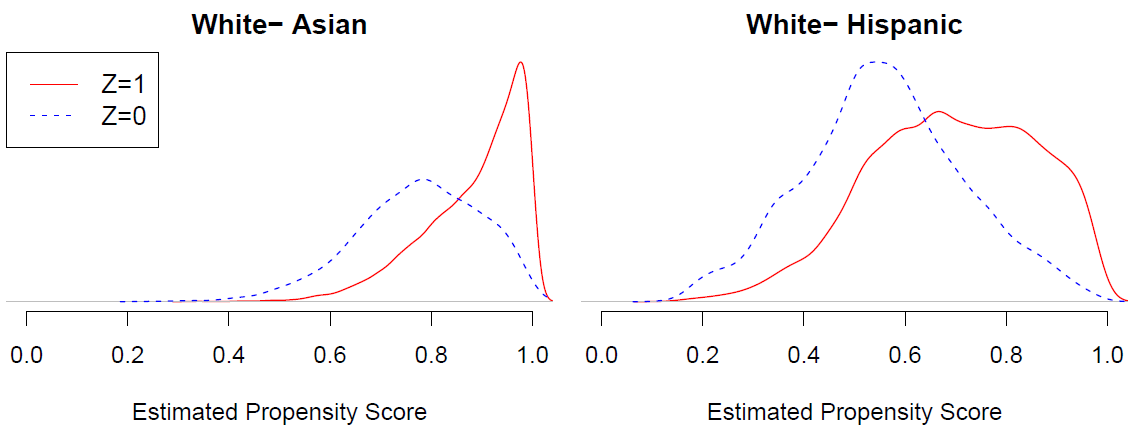}
\caption{Distribution of the estimated propensity scores for White-Asian and White-Hispanic comparisons in the MEPS data.}
\label{fig:dispWeights}
\end{figure}

There are 29 covariates (4 continuous, 25 binary), a mix of health indicators and demographic variables. We estimate the propensity scores via a logistic regression including all main effects. For each comparison $Z = 1$ for Whites and $Z =0$ for minority individuals. The distributions of the estimated propensity scores for the White-Asian and White-Hispanic comparisons are shown in Figure \ref{fig:dispWeights}. The White-Asian comparison has severe lack of overlap, with the largest Asian HT normalized weight being 0.32, meaning that one individual (out of 1522) accounts for almost one third of the total weights of Asians. This weight belongs to an Asian female with a very high BMI of 55.4 (the highest among Asians) and consequently a propensity score close to 1. In contrast, the largest overlap normalized weight is only 0.0008.

\begin{figure}[!ht]
\centering
\includegraphics[width=.99\textwidth]{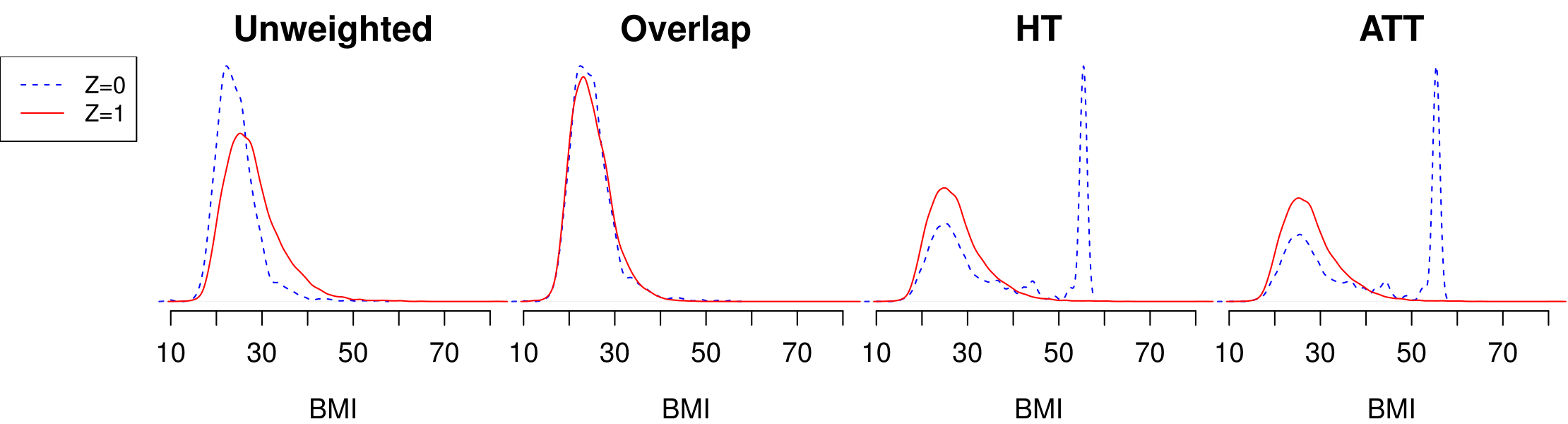}
\caption{Unweighted and weighted BMI distributions for the White and Asian groups in the MEPS data.}
\label{fig:bmi}
\end{figure}

Figure~\ref{fig:bmi} provides a closer look at the covariate BMI for the White-Asian comparison, showing the unweighted and weighted BMI distributions under each weighting scheme.  This illustrates the good balance achieved by the overlap weights, and the bad balance and extreme weight placed on the highest Asian BMI under the inverse probability weighting schemes.

\begin{figure}[!ht]
\centering
\includegraphics[width=.9\textwidth]{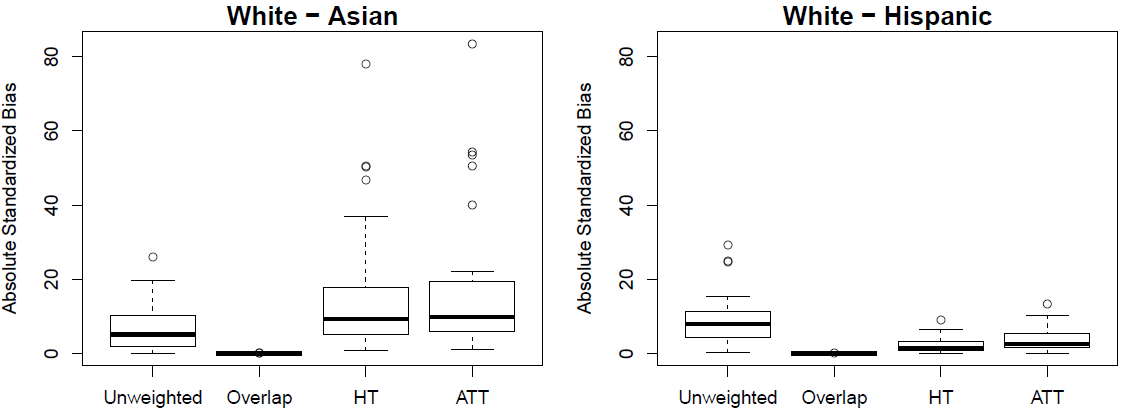}
\caption{Boxplots for the absolute standardized difference for covariates under each weighting method. }\label{fig:mepsbox}
\end{figure}

Figure~\ref{fig:mepsbox} shows boxplots of the ASB for all covariates under each weighting method. As expected from Theorem 3, the overlap weights always lead to perfect mean balance. For the White-Hispanic comparison, HT and ATT weighting each substantially improved mean balance compared to the unweighted data, although not as much as overlap weighting. For the White-Asian comparison, where there is serious lack of overlap in covariates, HT and ATT weighting results in worse covariate balance than no weighting at all, yielding very large differences in means for several covariates, including a difference of over 74 unweighted standard errors for the covariate BMI.

We estimate the weighted average controlled difference in health care expenditure between races using the estimator $\hat{\tau}_h$. The results with bootstrap standard errors (from 1000 bootstrap samples) appear in Table~\ref{tab:meps_est}. Estimates differ substantially across weighting methods, especially when there is a serious lack of overlap, as in the White-Asian comparison. For example, the average weighted difference in health care expenditure between Whites and Asians is estimated to be \$2772, \$1302, \$2458, \$2624 from the un-, overlap-, HT- and ATT-weighted methods, respectively.  The IPW weighting methods make the Asian disparity appear to be much greater than the Hispanic disparity, while this difference is negligible when using the overlap weights.  Overlap weighting focuses on the population where the Whites and minority groups have the most similar characteristics, and gives the smallest standard error in all the comparisons.

\begin{table}[!ht]
\centering
\renewcommand{\baselinestretch}{1.1}
\normalsize
\begin{tabular}{c|cccc}
\hline
 & Unweighted (se) & Overlap (se) & HT (se) & ATT (se) \\
\hline
White - Asian & 2772.0 ~ (225.1) & 1301.9 ~(220.0) & 2457.8~(575.7) & 2623.9~(633.8) \\
White - Hispanic & 2562.6 ~ (177.2)& 1292.0 ~(160.5) & 511.6 ~(358.6) & 50.8~(508.4) \\
\hline
\end{tabular}
\caption{Weighted mean differences in total health care expenditure (dollars) in 2009.}
\label{tab:meps_est}
\end{table}

The common practice of discarding or truncating units with propensity scores close to 0 or 1 can make estimates very dependent on the truncation cutoff.  We illustrate this by using HT weights after such truncation or discarding to estimate White - Hispanic disparities. These results in Table~\ref{tab:meps_truncate} show estimates changing more than two-fold based on the cutoff choice.  In contrast, the overlap weights avoid this artificial truncation or discarding, and rather continuously and automatically down-weight units with extreme propensity scores.

\begin{table}[ht]
\centering
\renewcommand{\baselinestretch}{1.1}
\normalsize
\begin{tabular}{c|cccc}
\hline
 &\multicolumn{4}{c}{Estimated Propensity Score Range Kept} \\
 & [0,1] & [0.01, 0.99] & [0.05, 0.95] & [0.1, 0.9] \\
\hline
Truncate & 512 & 510 & 1012 & 1347\\
Discard  & 512 & 446 & 1247 & 1166 \\
\hline
\end{tabular}
\caption{HT weighted mean differences (White - Hispanic) in total health care expenditure (dollars) in 2009 after truncating or discarding units with extreme propensity scores.}
\label{tab:meps_truncate}
\end{table}

\section{Discussion and Extensions} \label{sec:conclusion}

Covariate balance between comparison groups is central to causal and unconfounded descriptive studies. In this paper, we propose a unified framework for the class of weights---the balancing weights---that balance covariates. Several familiar types of weights, such as the HT and ATT weights, are special cases. Within this class, we advocate the overlap weights, which optimize the efficiency of comparisons by defining the population with the most overlap in the covariates between treatment groups. This weighting method is easy to implement with standard software and has been used in a number of applications. Though the overlap weights are statistically motivated, we argue that the corresponding target population and estimand are often of scientific or policy interest. Our method differs from several published methods for automated covariate balancing \cite[e.g.][]{Hainmueller12, Graham12,ImaiRatkovic14} that identify \emph{estimates of propensity scores (weights)} under certain criterion (e.g., moment conditions).  Rather, we identify a \emph{target population} that minimizes asymptotic conditional variance of the treatment effect without preliminary ``peeking" at the outcome data, as in \cite{Crump06}, and then identify additional properties, salutary in applications, of the corresponding weights.

The results of Section \ref{sec:largesample} readily extend under the same assumptions to optimizing a common target distribution $f(x)h(x)$ for comparisons of $J>2$ groups \citep{Imbens00,ImaiDVD04}. Suppose $e_j(x), j=1,\dots,J$ are conditional probabilities of assignment to group $j$, with $\sum e_j(x)=1$, and that the objective is to minimize the sum of the variances of weighted group means.  Then the balancing weights are $w_j(x)=h(x)/e_j(x)$, and under the conditions of Corollary 1 they are optimized for $h(x)\propto\left(\sum 1/e_j(x)\right)^{-1}$.  Extensions to (known) heteroscedasticity or to an unequally-weighted objective function are straightforward.  However, comparison of multiple groups allows consideration of a larger selection of propensity score model specifications than the two-group comparison, as well as different sets of comparisons of interest.  Heuristically, $h(\cdot)$ gives the most relative weight to the covariate regions in which {\em none} of the $e_j(\cdot)$ are close to zero.  With multiple groups this region of joint overlap might be small or nonexistent, even if all the pairwise overlaps are substantial.  Thus, the suitability of weighting to a common distribution depends on the specifics of covariate distributions, models, and scientific objectives of the analysis. Similar issues arise in matching of multiple groups.

Propensity-weighting methods can be readily extended to incorporate sampling weights \citep{DuGoff14}, which is arguably an advantage over the matching methods. In one common approach, the propensity score model $e_S(x)$ is estimated using sampling-weighted estimators, and then  observation $i$ with sampling weight $W_i$ and corresponding balancing weight $w_S(x_i)$ is weighted by $w_S(x_i)W_i$, where $w_S(x_i)=h(x_i)/e_s(x_i)$ for $Z_i=1$, $w_S(x_i)=h(x_i)/(1-e_s(x_i))$ for $Z_i=0$.  The weighted sample distribution is an unbiased and consistent estimator of the finite population distribution, and the estimation procedure (propensity score weighting) is consistent when applied to the finite population. On the other hand, the sampling-weighted estimator is likely to have increased variance if the combined weights $w_S(x_i)W_i$ are more variable than the propensity-score weights from the unweighted model $w_U(x_i)$. Furthermore, with weighted data the variance estimator and optimality arguments in Section \ref{sec:largesample} must be modified because the variance of the weighting estimator is no longer dependent only on sample size, but also on the weight distribution. Similar issues apply to the method of \cite{Zanutto06}, which combines unweighted subclassification modeling with weighted estimation of subclass means. A lively stream of current research aims to recover information from the weights without the inefficiency of the standard weighted analysis \citep{Zhen:Litt:infe:2005}; these approaches might be productively applied to propensity-score weighting.

\section*{Appendix}

In what follows we assume regularity conditions on $v_z$ and $\bE[Y(z)\mid X]$ necessary to make the integrals defined and convergent.

\noindent \textbf{Proof of Theorem 1.}
The  WATE for the population with density proportional to $f(x)h(x)$ with respect to base measure $\mu$ is defined as
\begin{eqnarray}
\tau_h &=&\int \tau(x) f(x)h(x) \mu(dx) \Big/\int f(x)h(x)\mu(dx) \nonumber \\
&=&\frac{\int \bE_{Y,Z|X} \left\{Y(1) Z[h(x)/e(x)]  - Y(0) (1-Z)[h(x)/(1-e(x))]\right\} f(x) \mu(dx)}{\int h(x) f(x) \mu(dx)} \nonumber
\end{eqnarray}
\begin{equation}
= \frac{\int \bE_{Y,Z|X} Y(1) Z[h(x)/e(x)] f(x)\mu(dx)}{\int \bE_{Z|X} Z[h(x)/e(x)]f(x) \mu(dx)}
- \frac{\int \bE_{Y,Z|X} Y_0 (1-Z)[h(x)/e(x)]f(x)\mu(dx)}{\int \bE_{Z|X} (1-Z)[h(x)/e(x)]f(x) \mu(dx)} \label{eq:tau}
\end{equation}
where $\tau(x)=\bE[Y(1)-Y(0)\mid X=x]$, and using the unconfoundedness assumption that $Y(1),Y(0)\perp Z\mid X$.  The terms of \eqref{eq:tau} can be read as expectations of weighted means of $Y(z)$ in samples drawn from the population with density $f(x)$, respectively for the strata with $z=1$ or $z=0$.  Replacing expectations by sample means, and substituting weight expressions from \eqref{weight}, we obtain the following estimator for the sample WATE:
\begin{equation}
\label{eq:sample}
\hat{\tau}_h=\frac{\sum_i Y_i(1)Z_iw_1(x_i)}{\sum_i Z_i w_1(x_i)}-
\frac{\sum_i Y_i(0)(1-Z_i)w_0(x_i)}{\sum_i (1-Z_i)w_0(x_i)}
\end{equation}
where each summation (divided by $n$) is an unbiased estimator of the corresponding integral in \eqref{eq:tau}; therefore by Slutsky's theorem $\hat{\tau}_h$ is a consistent estimator of $\tau_h$.

\noindent \textbf{Proof of Theorem 2.}
Conditional on the sample $\bX=\{x_1,\dots,x_n\}$ and $\bZ=\{z_1,\dots,z_n\}$, only $Y_i$ is random in \eqref{eq:sample}, so the variance of the estimator $\hat{\tau}_h$ is
\begin{eqnarray}
\bV[\hat{\tau}_h\mid \bX, z] &=&\frac{\sum_i v_1(x_i)z_iw_1(x_i)^2}{\left[\sum_i z_i w_1(x_i)\right]^2}+
\frac{\sum_i v_0(x_i)(1-z_i)w_0(x_i)^2}{\left[\sum_i (1-z_i)w_0(x_i)\right]^2} \nonumber \\
&=& \frac{\sum_i v_1(x_i)[z_i/e(x_i)][h(x_i)^2/e(x_i)]}{\left\{\sum_i [z_i/e(x_i)]h(x_i)\right\}^2}+
\frac{\sum_i v_0(x_i)[(1-z_i)/(1-e(x_i))][h(x_i)^2/(1-e(x_i))]}{\left\{\sum_i [(1-z_i)/(1-e(x_i))]h(x_i)\right\}^2}.  \nonumber
\end{eqnarray}
Averaging the above first over the distribution of $\bZ$ (using $\bE [Z_i/e(x_i)]=\bE [(1-Z_i)(1-e(x_i))]=1$), and then over the distribution of $\bX$, and again applying Slutsky's theorem, we have
$$
n\cdot\bE_x\bV[\hat{\tau}_h\mid \bX] \rightarrow \int \left(\frac{v_1(x)}{e(x)}+\frac{v_0(x)}{1-e(x)}\right)h(x)^2 f(x)
 \mu(dx) \Big/\left(\int f(x)h(x)\mu(dx)\right)^2.
$$

\noindent \textbf{Proof of Corollary 1.}
For simplicity, we use $\bE[\cdot]$ to denote $\int\cdot f(x)\mu(dx)$. According to the Cauchy-Schwarz inequality, we have
\begin{eqnarray*}
\left[  \bE\left\{ h(x)\right\}\right]^2 &=&  \left[  \bE\left\{\frac{h(x)}{\sqrt{e(x)(1-e(x))}} \sqrt{ e(x)(1-e(x))}\right\}\right]^2 \\
 &\leq & \bE\left\{\frac{h^2(x)}{e(x)(1-e(x))}\right\} \bE\left[e(x)(1-e(x))\right],
\end{eqnarray*}
and the equality is attained when $\frac{h(x)}{\sqrt{e(x)(1-e(x))}} \propto \sqrt{e(x)(1-e(x))}$, that is, when $h(x)\propto e(x)(1-e(x))$. Corollary 1 follows directly from applying the above to the right hand side of \eqref{simplevar}.

\noindent \textbf{Proof of Theorem 3.}  The score functions of the logistic propensity score model, $\mbox{logit}\{e(x_i)\}=\beta_0+x_i\beta'$ with $\beta=(\beta_1,...,\beta_K)$, are:
\begin{equation*}
\frac{\partial \log L}{\partial \beta_k} = \sum_i x_{ik} (Z_i -\eh), \quad \mbox{for} ~~ k=0, 1,\dots, K,
\end{equation*}
where $x_{0k}\equiv 1$ and $\eh=\left[1+\exp(-x_i \beta')\right]^{-1}$.
Equating to 0 and solving, the MLE $\hat\beta$ satisfies
\[\sum Z_i = \sum \hat{e}_i, \quad \mbox{and} \quad  \sum x_{ik} Z_i = \sum x_{ik} \hat{e}_i.\]
It follows that
\begin{eqnarray*}
\sum_i Z_i(1-\hat{e}_i)&=& \sum \hat{e}_i - \sum_i Z_i\hat{e}_i=\sum \hat{e}_i(1-Z_i), \\
\sum_i x_{ik}Z_i(1-\hat{e}_i)&=& \sum x_{ik} \hat{e}_i - \sum_i x_{ik}Z_i\hat{e}_i=\sum x_{ik} \hat{e}_i(1-Z_i), \quad \mbox{for} ~~ k=1,\dots, K.
\end{eqnarray*}
Therefore, for any $k=1,\dots,K$, we have
\[\frac{\sum_i x_{ik}Z_i(1-\hat{e}_i)}{\sum_i Z_i(1-\hat{e}_i)}
=\frac{\sum_i x_{ik}(1-Z_i)\hat{e}_i}{\sum_i (1-Z_i)\hat{e}_i}. \quad  \]

\bibliographystyle{jasa3}
\bibliography{psweight}

\end{document}